\documentclass[12pt]{article}
\usepackage{amssymb}
\usepackage{amsmath}

\begin{document}

\begin{center}
\vspace{1cm}{\Large {\bf 10d N=1 Massless BPS supermultiplets}}

\vspace{1cm} {\bf H. Mkrtchyan} \footnote{ E-mail:hike@r.am} and
{\bf R. Mkrtchyan} \footnote{ E-mail: mrl@r.am} \vspace{1cm}

\vspace{1cm}

 {\it Yerevan Physics Institute, }{\it Theoretical Physics Department}

{\it Alikhanian Br. St.2, Yerevan 375036, Armenia}
\end{center}

\vspace{1cm}
\begin{abstract}

We consider d=10 N=1 supersymmetry algebra with maximal number of
tensor charges Z and introduce a class of orbits of Z, invariant
w.r.t. the $T_8$ subgroup of massless particles' little group
$T_8\ltimes SO(8)$. For that class of orbits we classify all
possible orbits and little groups, which appear to be semidirect
products of $T_8\ltimes SO(k_1)\times ... SO(k_n)$ form, with
$k_1+...+k_n=8$, where compact factor is  embedded  into SO(8) by
triality map. We define actions of little groups on supercharge Q
and construct corresponding supermultiplets. In some particular
cases we show the existence of supermultiplets with both Fermi and
Bose sectors consisting of the same representations of tensorial
Poincare. In addition, complete classification of supermultiplets
(not restricted to $T_8$-invariant orbits) with rank-2 matrix of
supersymmetry charges anticommutator, is given.
\end{abstract}

\renewcommand{\thefootnote}{\arabic{footnote}}
\setcounter{footnote}0 {\smallskip \pagebreak }

\section{Introduction}

The basic blocks of a theory with a symmetry are the unitary
irreducible representations of it's symmetry group. If symmetry is
unbroken, then states of Gilbert space would classify according to
unitary irreps of symmetry. If we have a Lagrangian description of
a theory, it is usually either linear, or non-linear
representation of symmetry. In the relativistic theory, invariant
w.r.t. the Poincare symmetry, corresponding unitary irreps are
called particles, with different masses and spins.

In modern superbranes theories (superstring/M-theory) the Poincare
group is substituted by a new one, which includes in addition to
generators of translations in the space time new tensorial
generators \cite{TowAsk} (which can be considered as generators of
translations in a generalized space-time \cite{Ask}, \cite{Band2},
\cite{Vas}, \cite{Man}). The characteristic relation is
anticommutator of supercharges, given below for 10d N=1:

\begin{eqnarray}\label{eq1.9}
\left\{ Q_\alpha,Q_\beta \right\} &=& Z_{\alpha\beta}\\
&=&(\Gamma ^{\mu}C)_{\alpha\beta}P_{\nu}+
(\Gamma ^{\mu\nu\lambda\rho\sigma}C)_{\alpha\beta}Z_{\mu\nu\lambda\rho\sigma}^{+}\nonumber\\
  \mu, \nu,...
&=&0,1,2,..9. \nonumber
\end{eqnarray}

Now well-known states in supergravity/superstrings theories are
brane states, which appear in different ways, e.g. as solutions of
classical equations of motion, but, finally, are just various
representations of the super-Poincare algebra (\ref{eq1.9}) (see,
e.g. \cite{Tow}. Study of the representations of superalgebra
(\ref{eq1.9}) is the main aim of the present paper. Particularly,
we are interested in construction of BPS multiplets with
additional restrictions on little groups of representations. Most
interesting among such restrictions seems to be those which lead
to zero values of as many invariants as possible, constructed from
the central charges Z. Very often such representations play
fundamental role in the theory, as e.g. massless particles with
different spins, if we consider the well known physical examples
of four-dimensional theory - Yang-Mills, gravity, etc. Some
results on 10d N=1 massive BPS supermultiplets are obtained in
\cite{Hik}.

Explicit discussion  of some BPS supermultiplets of tensorial
super-Poincare algebras from the group theory's  point of view is
given in \cite{mrl2}, \cite{mrl3}. The feature, which can be
considered as the most unusual, in comparison with usual Poincare
supermultiplets, is the possibility of existence of
supermultiplets in which fermionic and bosonic members are in the
same representation of bosonic subalgebra (tensorial Poincare) of
superalgebra. Remind, that in usual supersymmetry the members of a
supermultiplet differ by their spin, those with half-integer spins
have Fermi statistics, those with integer spin - Bose one, in
agreement with spin-statistics theorem. Supercharge, according to
its fermionic nature, transforms half-integer spins into integer
ones, and vice-versa. For tensorial super-Poincare, in the most
simple preon case, i.e. for orbits with rank-one Z matrix,
supermultiplet consists of two members, both in the same
representation of the little group, supercharge has zero spin, and
simply transforms one state into another, without rotation in the
space of representation. Particularly, in the simplest case these
representations can be singlet, so supermultiplet consists of
exactly two states, one of them is assigned a Fermi statistics,
and the other - Bose one. We shall see below the generalizations
of this example, when again, both fermionic and bosonic members of
supermultiplet are in the same representation of little group, but
supercharge is non-trivially acting inside that representations.
Alltogether,  these features mean that some generalization of spin
statistics theorem is necessary for these new space-time symmetry
algebras (\ref{eq1.9}), see \cite{mrl2}, \cite{mrl3}, \cite{mrl4}.
One possibility for super-preon representation is to lift that to
the super-singleton representation of OSp group (for a review, see
e.g. \cite{S}). Then two members of preon supermultiplet become
two different singleton representations of Sp group, with spin
zero and 1/2, respectively. It is not clear, whether such a
phenomena appears in other cases, namely, whether other, similar
to preon, representations can be lifted to representations of OSp
group, although a necessary condition of masslessness is
fulfilled.

In Section 2 we present the definition of a class of BPS
supermultiplets, to which present paper is devoted. It is
distinguished by existence of $T_8$ subgroup in corresponding
(semidirect product) little group. We calculate possible little
groups, presenting an examples of orbits for main cases. Section 3
is devoted to construction of supermultiplets for main cases, in
that section we show the preon-type supermultiplets with fermions
and bosons in the same representations of the little group.
Complete classification, including a massive case, for a rank-2 Z
matrix is given in Section 4. In the Section 5 we give a
definition of specific basis of gamma-matrices, used in
calculations and in presentation of results.

\section{Little groups}
Main object of our considerations is d = 10 N = 1 susy algebra
with central charges, given by (\ref{eq1.9}). We shall use a
specific (standard) representation of gamma-matrices, described at
the end of this paper. Sometimes it will be easier to formulate
and prove statements using that specific representation.

The unitary irreps of Poincare, super-Poincare and their tensorial
generalization (\cite{TowAsk}), with tensorial charges added on an
equal footing to energy-momentum vector, can be constructed by the
method of induced representations \cite{Wig}. That process
includes choosing of specific orbit of the action of Lorentz group
on the space of charges Z, calculation of stability group (little
group) of some arbitrary point on that orbit, construction of an
unitary irrep of little group, extending it to an unitary irrep of
superalgebra (\ref{eq1.9}), and finally induction of that irrep on
the whole super-Poincare group.  The first step in this procedure,
i.e. the choice of an orbit in the space of Z-matrix, can be
started e.g. from choosing values for the invariants constructed
from Z.

A massless particle's momenta $p_\mu$ is invariant w.r.t. the
$T_8\ltimes SO(8) $ subgroup of Lorentz SO(1,9). We choose a
specific form $p_\mu=(1,0,...1)$, then $t_8$ subalgebra is a 10 by
10 matrix with non-zero elements at upper row with zeros at first
and last positions, first column is recovered by antisymmetry,
last column is equal to the first one with minus sign, bottom row
follows from antisymmetry. We put a restriction on Z matrix to be
invariant w.r.t. this $T_8$ subgroup. In our representation that
implies that non-zero elements of Z are concentrated in left upper
8x8 corner. Similarly, if we require an invariance of some general
spinor w.r.t. $T_8$ group, the last 8 components of that spinor
will be zero, upper ones remain arbitrary. In invariant terms: an
SO(1,9) Majorana-Weyl 16-component spinor decomposes into two
SO(8) 8-component spinors (upper and lower halves, in our basis)
of the subgroup $SO(8)\ltimes T_8$, $T_8$ transforms lower spinor
into upper, leaving upper one unchanged, so $T_8$ invariant spinor
has non-zero upper half, only. Important statement is that there
exists Lorentz transformation, which acts on that $T_8$ invariant
spinors (and on Z, also) as scale transformation, simply
multiplying them by a numerical factor. Namely, generator of
spinor's rotation  in (09) plane is given by $(1/4)[\Gamma^{0},
\Gamma^{9}]$, which in our representation is diagonal matrix with
1 on first 8 positions and -1 on next 8 ones (and 1 on next 8 and
-1 on remaining 8 positions). So it is proportional to an identity
matrix for our restricted spinors and on $Z$, so corresponding
group element is acting as multiplication by number. Important
consequence is that all polynomial Lorentz invariants constructed
from our $T_8$ invariant $Z$ should be zero. Discrete invariants,
such as rank, can be non-trivial.

One can reverse the way of thinking, and instead of asking a
question what is the little group of a given Z matrix, can choose
some subgroup of Lorentz group and ask what are the Z matrices
with a given subgroup as a little group.

Next we would like to make a statement on the little groups of Z's
with little groups of the form  $T_8\ltimes G$. The problem is to
define a little groups of symmetric second rank $8\times 8$ matrix
with indexes from spinor Majorana-Weyl representation. Applying an
appropriate triality transformation, we can transform this problem
into the problem of finding little groups of symmetric second rank
tensor with indices from vectorial representation. This problem
can be solved by diagonalizing this matrix, with $k_1$ eigenvalues
$a_1$, $k_2$ eigenvalues $a_2$,etc. Then one can easily understand
that little group of such a matrix is $SO(k_1)\otimes SO(k_2)
\otimes ...$. Later on we shall define the action of little group
on supercharges and, finally, will construct corresponding
supermultiplets.

For construction of specific examples of all these possible little
groups, we introduce the following basis in the spinor space:

\begin{eqnarray}
p_\alpha ^{i} = \begin{array}{*{20}c}
   1 & i & 0 & 0 & 0 & 0 & 0 & 0  \\
   0 & 0 & 1 & i & 0 & 0 & 0 & 0  \\
   0 & 0 & 0 & 0 & 1 & i & 0 & 0  \\
   0 & 0 & 0 & 0 & 0 & 0 & 1 & i  \\
   0 & 0 & { - 1} & i & 0 & 0 & 0 & 0  \\
   1 & { - i} & 0 & 0 & 0 & 0 & 0 & 0  \\
   0 & 0 & 0 & 0 & 0 & 0 & 1 & { - i}  \\
   0 & 0 & 0 & 0 & 0 & { - 1} & i & 0  \\
\end{array}
\end{eqnarray}

Lower index runs over rows, and corresponds to components of
spinor, upper index enumerates different spinors of basis. Let's
consider the following form of the matrix Z:

\begin{eqnarray}\label{z}
Z_{\alpha\beta}= \sum\limits_{i = 1}^8 {a_i p_\alpha ^i p_\beta
^i}
\end{eqnarray}

The eigenvalue problem we should consider is
$det(Z_{\alpha\beta}-x B_{\alpha\beta})=0$, or
$det(Z_{\alpha\beta}B^{\beta\gamma}-x)=0$, where $B_{\alpha\beta},
BB^*=1$ is a 8x8 restriction of a matrix of complex conjugation of
gamma-matrices in ten-dimensional Minkowski space. Note that
according to susy relation (\ref{eq1.9}) the eigenvalues of
$(Z_{\alpha\beta}B^{\beta\gamma})$ should be non-negative.
Corresponding eigenvalues of matrix (\ref{z}) are $2a_1, 2a_2,
...$, so these numbers should be taken non-negative. Without loss
of generality we can assume that first $k_1$ elements of set
$(a_1,a_2,...)$ are equal to each other, then next $k_2$ elements
are equal, etc., $\sum k_i=8$. Then the little group of Z is
$T_8\ltimes SO(k_1)\otimes SO(k_2)\otimes ....$. But this is not
enough, one has to specify the embedding of compact factor into $
SO(8)$. That embedding is given by inverse of above mentioned
triality transformation, under which vector representation is
going back to spinor one. Already at this point we can mention
that this result coincides with result in preon case \cite{mrl4},
when non-zero is $a_1$ only, other $a$-s are equal to zero, and
the little group is $SO(1)\otimes SO(7)\sim SO(7)$, spinorially
embedded into $SO(8)$, i.e. so(7) is embedded into so(8) in its
spinorial 8-dimensional irreducible (Majorana) representation.
Generally, the $SO(k_1)\otimes SO(k_2)\otimes ....$, according to
above mentioned is embedded in the spinorial representation. Below
we show these embeddings explicitly in a few main cases.

$SO(7)\otimes SO(1)$: the $so(7)$ subalgebra of $so(8)$ is (here
and below we use a given representation of gamma-matrices, see
"Notations")

\begin{eqnarray}\label{so7}
\begin{pmatrix}
  0 & -c_{1} & -c_{2} & -c_{3} & -c_{4} & -c_{5} & -c_{6} & -c_{7}\\
 c_{1} & 0 & w_{34} & w_{35} & w_{36} & w_{37} & w_{38} & w_{39}\\
 c_{2} & -w_{34} & 0 & w_{45} & w_{46} & w_{47} & w_{48} & w_{49}\\
 c_{3} & -w_{35} & -w_{45} & 0 & w_{56} & w_{57} & w_{58} & w_{59}\\
 c_{4} & -w_{36} & -w_{46} & -w_{56} & 0 & w_{67} & w_{68} & w_{69}\\
c_{5} & -w_{37} & -w_{47} & -w_{57} & -w_{67} & 0 & w_{78} & w_{79}\\
 c_{6} & -w_{38} & -w_{48} & -w_{58} & -w_{68} & -w_{78} & 0 & w_{89}\\
c_{7} & -w_{39} & -w_{49} & -w_{59} & -w_{69} & -w_{79} & -w_{89} & 0 &\\
  \label{10L}
\end{pmatrix}
\end{eqnarray}

where $c_1=-w_{46} - w_{57} + w_{89}, c_2=w_{36} + w_{58} +
w_{79}, c_3=w_{37} - w_{48} - w_{69}, c_4=-w_{34} + w_{59} -
w_{78}, c_5=-w_{35} - w_{49} + w_{68}, c_6=-w_{39} + w_{45} -
w_{67}, c_7=w_{38} + w_{47} - w_{56} $.

The algebra of $SO(6)\otimes SO(2)$ little group is represented
by:
\begin{eqnarray}\label{so62}
\begin{array}{*{20}c}
   {\rm{0}} & {{\rm{ - s}}_{{\rm{78}}} } & {{\rm{s}}_{68} } & {{\rm{ - s}}_{58} } & {{\rm{s}}_{48} } & {{\rm{ - s}}_{38} } & {{\rm{s}}_{28} } & {{\rm{s}}_{18} }  \\
   {{\rm{s}}_{{\rm{78}}} } & {\rm{0}} & {{\rm{s}}_{67} } & {{\rm{ - s}}_{57} } & {{\rm{s}}_{{\rm{47}}} } & {{\rm{ - s}}_{{\rm{37}}} } & {{\rm{s}}_{{\rm{27}}} } & {{\rm{s}}_{{\rm{28}}} }  \\
   {{\rm{ - s}}_{68} } & {{\rm{ - s}}_{67} } & {\rm{0}} & {{\rm{s}}_{{\rm{56}}} } & {{\rm{ - s}}_{{\rm{46}}} } & {{\rm{s}}_{{\rm{36}}} } & {{\rm{s}}_{{\rm{37}}} } & {{\rm{s}}_{{\rm{38}}} }  \\
   {{\rm{s}}_{58} } & {{\rm{s}}_{57} } & {{\rm{ - s}}_{{\rm{56}}} } & {\rm{0}} & {{\rm{s}}_{{\rm{45}}} } & {{\rm{s}}_{{\rm{46}}} } & {{\rm{s}}_{{\rm{47}}} } & {{\rm{s}}_{{\rm{48}}} }  \\
   {{\rm{ - s}}_{48} } & {{\rm{ - s}}_{{\rm{47}}} } & {{\rm{s}}_{{\rm{46}}} } & {{\rm{ - s}}_{{\rm{45}}} } & {\rm{0}} & {{\rm{s}}_{{\rm{56}}} } & {{\rm{s}}_{{\rm{57}}} } & {{\rm{s}}_{{\rm{58}}} }  \\
   {{\rm{s}}_{38} } & {{\rm{s}}_{{\rm{37}}} } & {{\rm{ - s}}_{{\rm{36}}} } & {{\rm{ - s}}_{{\rm{46}}} } & {{\rm{ - s}}_{{\rm{56}}} } & {\rm{0}} & {{\rm{s}}_{{\rm{67}}} } & {{\rm{s}}_{{\rm{68}}} }  \\
   {{\rm{ - s}}_{28} } & {{\rm{ - s}}_{{\rm{27}}} } & {{\rm{ - s}}_{{\rm{37}}} } & {{\rm{ - s}}_{{\rm{47}}} } & {{\rm{ - s}}_{{\rm{57}}} } & {{\rm{ - s}}_{{\rm{67}}} } & {\rm{0}} & {{\rm{s}}_{{\rm{78}}} }  \\
   {{\rm{ - s}}_{18} } & {{\rm{ - s}}_{{\rm{28}}} } & {{\rm{ - s}}_{{\rm{38}}} } & {{\rm{ - s}}_{{\rm{48}}} } & {{\rm{ - s}}_{{\rm{67}}} } & {{\rm{ - s}}_{{\rm{68}}} } & {{\rm{ - s}}_{{\rm{78}}} } & {\rm{0}}  \\
\end{array}\label{so62}
\end{eqnarray}
This matrix coincides with that of $u(4)\sim so(6)+ so(2)$, as can
be seen in the following way. Take general $u(4)$ $4\times 4$
anti-Hermitian matrix $Z=X+iY$ with real antisymmetric X and real
symmetric Y matrices, then represent it in a real basis, as:

\begin{eqnarray}
\begin{array}{*{20}c}
Z_R = \begin{array}{*{20}c}
   X & { - Y}  \\
   Y & X  \\
\end{array}
\end{array}
\end{eqnarray}
and finally make a transformation

\begin{eqnarray}
Z_R\rightarrow S^{-1}Z_R S
\end{eqnarray}
with
\begin{eqnarray}
 S = \begin{array}{*{20}c}
   0 & 0 & 0 & 0 & 1 & 0 & 0 & 0  \\
   0 & 0 & 0 & 0 & 0 & 1 & 0 & 0  \\
   0 & 0 & 0 & 0 & 0 & 0 & 1 & 0  \\
   0 & 0 & 0 & 0 & 0 & 0 & 0 & 1  \\
   0 & 0 & 0 & 1 & 0 & 0 & 0 & 0  \\
   0 & 0 & { - 1} & 0 & 0 & 0 & 0 & 0  \\
   0 & 1 & 0 & 0 & 0 & 0 & 0 & 0  \\
   1 & 0 & 0 & 0 & 0 & 0 & 0 & 0  \\
\end{array}
\end{eqnarray}

which will give exactly the matrix (\ref{so62}). Note that
fundamental 4-dimensional representation of $su(4)$ is a spinorial
(Weyl) representation of so(6).

 $SO(5)\otimes SO(3)$, the corresponding matrix is:

\begin{eqnarray}\label{so53}
\begin{array}{*{20}c}
   {\rm{0}} & {\rm{.}} & . & . & . & {\rm{.}} & . & .  \\
   {{\rm{ - s}}_{{\rm{35}}} {\rm{ - s}}_{{\rm{46}}} {\rm{ - s}}_{{\rm{78}}} } & {\rm{0}} & . & . & {\rm{.}} & {\rm{.}} & {\rm{.}} & {\rm{.}}  \\
   {{\rm{ - s}}_{68} } & {{\rm{ - s}}_{67} } & {\rm{0}} & {\rm{.}} & {\rm{.}} & {\rm{.}} & {\rm{.}} & {\rm{.}}  \\
   {{\rm{s}}_{37} } & {{\rm{ - s}}_{38} } & {{\rm{ - s}}_{{\rm{34}}} } & {\rm{0}} & {\rm{.}} & {\rm{.}} & {\rm{.}} & {\rm{.}}  \\
   {{\rm{ - s}}_{67} } & {{\rm{ - s}}_{{\rm{68}}} } & {{\rm{ - s}}_{{\rm{35}}} } & {{\rm{ - s}}_{{\rm{45}}} } & {\rm{0}} & {\rm{.}} & {\rm{.}} & {\rm{.}}  \\
   {{\rm{s}}_{38} } & {{\rm{s}}_{{\rm{37}}} } & {{\rm{ - s}}_{{\rm{36}}} } & {{\rm{ - s}}_{{\rm{46}}} } & {{\rm{ - s}}_{{\rm{56}}} } & {\rm{0}} & {\rm{.}} & {\rm{.}}  \\
   {{\rm{ - s}}_{{\rm{28}}} {\rm{ + s}}_{{\rm{34}}} {\rm{ - s}}_{{\rm{56}}} } & {{\rm{ - s}}_{{\rm{27}}} } & {{\rm{ - s}}_{{\rm{37}}} } & {{\rm{ - s}}_{{\rm{68}}} } & {{\rm{s}}_{{\rm{38}}} } & {{\rm{ - s}}_{{\rm{67}}} } & {\rm{0}} & {\rm{.}}  \\
   {{\rm{s}}_{{\rm{27}}} {\rm{ + s}}_{{\rm{36}}} {\rm{ - s}}_{{\rm{45}}} } & {{\rm{ - s}}_{{\rm{28}}} } & {{\rm{ - s}}_{{\rm{38}}} } & {{\rm{s}}_{{\rm{67}}} } & {{\rm{ - s}}_{{\rm{37}}} } & {{\rm{ - s}}_{{\rm{68}}} } & {{\rm{ - s}}_{{\rm{78}}} } & {\rm{0}}  \\
\end{array}
\end{eqnarray}
(here and below the omitted elements of matrices can be recovered
by antisymmetry).

This is a spinorial representation of $SO(5)\otimes SO(3)$, which
can be obtained by embedding this algebra into so(8), and taking
spinorial representation of the latter.

$SO(4)\otimes SO(4)$

The matrix is:

\begin{eqnarray}\label{so44}
\begin{array}{*{20}c}
   {\rm{0}} & {\rm{.}} & . & . & . & {\rm{.}} & . & .  \\
   {{\rm{ - s}}_{{\rm{12}}} } & {\rm{0}} & . & . & {\rm{.}} & {\rm{.}} & {\rm{.}} & {\rm{.}}  \\
   0 & 0 & {\rm{0}} & {\rm{.}} & {\rm{.}} & {\rm{.}} & {\rm{.}} & {\rm{.}}  \\
   0 & 0 & {{\rm{ - s}}_{{\rm{34}}} } & {\rm{0}} & {\rm{.}} & {\rm{.}} & {\rm{.}} & {\rm{.}}  \\
   0 & {\rm{0}} & {{\rm{ - s}}_{{\rm{35}}} } & {{\rm{ - s}}_{{\rm{45}}} } & {\rm{0}} & {\rm{.}} & {\rm{.}} & {\rm{.}}  \\
   0 & {\rm{0}} & {{\rm{ - s}}_{{\rm{36}}} } & {{\rm{ - s}}_{{\rm{46}}} } & {{\rm{ - s}}_{{\rm{56}}} } & {\rm{0}} & {\rm{.}} & {\rm{.}}  \\
   {{\rm{ - s}}_{{\rm{17}}} } & {{\rm{ - s}}_{{\rm{27}}} } & {\rm{0}} & {\rm{0}} & {\rm{0}} & {\rm{0}} & {\rm{0}} & {\rm{.}}  \\
   {{\rm{ - s}}_{{\rm{18}}} } & {{\rm{ - s}}_{{\rm{28}}} } & {\rm{0}} & {\rm{0}} & {\rm{0}} & {\rm{0}} & {{\rm{ - s}}_{{\rm{78}}} } & {\rm{0}}  \\
\end{array}
\end{eqnarray}

which seems to be a vectorial embedding of algebra of
$SO(4)\otimes SO(4)$ into so(8). Actually, if we have one so(4),
we shall obtain a sum of spinor representations of its two $so(3)$
factors:

\begin{eqnarray}\label{so4}
\begin{array}{*{20}c}
   {\rm{0}} & {\rm{.}} & . & . & . & {\rm{.}} & . & .  \\
   {{\rm{s}}_{{\rm{78}}} } & {\rm{0}} & . & . & {\rm{.}} & {\rm{.}} & {\rm{.}} & {\rm{.}}  \\
   0 & 0 & {\rm{0}} & {\rm{.}} & {\rm{.}} & {\rm{.}} & {\rm{.}} & {\rm{.}}  \\
   0 & 0 & {{\rm{ - s}}_{{\rm{56}}} } & {\rm{0}} & {\rm{.}} & {\rm{.}} & {\rm{.}} & {\rm{.}}  \\
   0 & {\rm{0}} & {{\rm{s}}_{{\rm{46}}} } & {{\rm{ - s}}_{{\rm{45}}} } & {\rm{0}} & {\rm{.}} & {\rm{.}} & {\rm{.}}  \\
   0 & {\rm{0}} & {{\rm{ - s}}_{{\rm{45}}} } & {{\rm{ - s}}_{{\rm{46}}} } & {{\rm{ - s}}_{{\rm{56}}} } & {\rm{0}} & {\rm{.}} & {\rm{.}}  \\
   {{\rm{ - s}}_{{\rm{28}}} } & {{\rm{ - s}}_{{\rm{27}}} } & {\rm{0}} & {\rm{0}} & {\rm{0}} & {\rm{0}} & {\rm{0}} & {\rm{.}}  \\
   {{\rm{s}}_{{\rm{27}}} } & {{\rm{ - s}}_{{\rm{28}}} } & {\rm{0}} & {\rm{0}} & {\rm{0}} & {\rm{0}} & {{\rm{ - s}}_{{\rm{78}}} } & {\rm{0}}  \\
\end{array}
\end{eqnarray}

Moreover, for the case of $SO(3)$ little group, the corresponding
embedding of so(3) into so(8) is:
\begin{eqnarray}\label{so3}
\begin{array}{*{20}c}
   {\rm{0}} & {\rm{.}} & . & . & . & {\rm{.}} & . & .  \\
   {{\rm{s}}_{{\rm{78}}} } & {\rm{0}} & . & . & {\rm{.}} & {\rm{.}} & {\rm{.}} & {\rm{.}}  \\
   0 & 0 & {\rm{0}} & {\rm{.}} & {\rm{.}} & {\rm{.}} & {\rm{.}} & {\rm{.}}  \\
   0 & 0 & {{\rm{ - s}}_{{\rm{56}}} } & {\rm{0}} & {\rm{.}} & {\rm{.}} & {\rm{.}} & {\rm{.}}  \\
   0 & {\rm{0}} & {{\rm{ - s}}_{{\rm{78}}} } & {{\rm{ - s}}_{{\rm{45}}} } & {\rm{0}} & {\rm{.}} & {\rm{.}} & {\rm{.}}  \\
   0 & {\rm{0}} & {{\rm{ - s}}_{{\rm{45}}} } & {{\rm{s}}_{{\rm{78}}} } & {{\rm{ - s}}_{{\rm{56}}} } & {\rm{0}} & {\rm{.}} & {\rm{.}}  \\
   {{\rm{ - s}}_{{\rm{56}}} } & {{\rm{ - s}}_{{\rm{45}}} } & {\rm{0}} & {\rm{0}} & {\rm{0}} & {\rm{0}} & {\rm{0}} & {\rm{.}}  \\
   {{\rm{s}}_{{\rm{45}}} } & {{\rm{ - s}}_{{\rm{56}}} } & {\rm{0}} & {\rm{0}} & {\rm{0}} & {\rm{0}} & {{\rm{ - s}}_{{\rm{78}}} } & {\rm{0}}  \\
\end{array}
\end{eqnarray}
which is the sum of two mutually conjugated spinor representations
of so(3), in a real basis.

\section{Supermultiplets, or Action of Little Group on Supercharge $Q$}

For construction of supermultiplet by the method of induced
representations, we have, particularly, to choose an orbit and
construct a representation of susy relation (\ref{eq1.9}) on that
orbit.This last step can be done either directly or by
representing (\ref{eq1.9}) as an algebra of (fermionic)
creation-annihilation operators, choosing the vacuum in a definite
representation of little group, then acting by creation operators.
Finally, knowing the action of little group on supercharge, one
can define the "spin content" of supermultiplet, i.e. what
representations of little group with what multiplicity are
contained in the constructed supermultiplet. Usually one is
interested in a minimal supermultiplet, with smallest "spins"
(smallest representations of little groups). We shall present a
supermultiplets for a few cases, with characteristic features.

For a Z-matrix (\ref{z}) of previous Section, $Q_\alpha$ has a
form:
\begin{eqnarray}
Q_\alpha=p_\alpha^i q_i \sqrt {a_i }\label{Qq}
\end{eqnarray}
We can satisfy susy commutation relations by choosing
(anti)commutation relations for $q_i$ as:
\begin{eqnarray}
 \{ q_i,q_j \}=  \delta_{ij}
\end{eqnarray}
and Majorana property of Q leads to the hermiticity  of $q_i$,
since $p_\alpha^i$ are Majorana spinors. So, for those $i$ with
$a_i=0$ corresponding terms are absent in (\ref{Qq}).

Next we turn to construction of specific examples of
supermultiplets.

Well-known example is the case when all $ a_i$-s are equal (to 1,
e.g.). Then invariance group is SO(8), all $Q_\alpha$-s ($q^i$-s)
are non-zero and transform as spinors. This is well-known case of
massless particle (and string's zero modes), and quantization,
(see e.g. \cite{GSW}) leads to $8_s+8_v=16$-dimensional space  of
spinors and vectors, fermions and bosons' respectively, with $q^i$
acting in this space as gamma-matrices for 8 Euclidean dimensions.
This representation can be accepted as "universal" one, in a sense
that it can be used for representing all other cases of little
group. Of course, the point is, that in some cases, with e.g. some
number of eigenvalues of Z equal to zero, that representation will
not be a minimal and/or irreducible one.

Next consider the preon case \cite{mrl3}, \cite{mrl4} when $a_1$
is non-zero, all other $a_i$ -s are equal to zero, and hence equal
to each other.

In this case little group is not acting on the Q at all, since
according to \ref{Qq} it is proportional to $\lambda_\alpha$,
which is stable under SO(7), since Z is stable. $q_1$ is
satisfying the anticommutation relation

\begin{eqnarray}
\{ q_1,q_1 \}=1
\end{eqnarray}
and in principle can be represented by a number. But this will
violate its fermionic nature, since it has to change statistics,
so the minimal representation is made possible with two by two
matrix:

\begin{eqnarray}
q_1  = \left| {\begin{array}{*{20}c}
   0 & 1  \\
   1 & 0  \\
\end{array}} \right|
\end{eqnarray}

So, supermultiplet has two members, both in the same
representation of tensorial Poincare, since supercharge Q,
transforming one into another, is neutral w.r.t. the little group
and correspondingly doesn't change its representation.

Next let's consider the following case with SO(3) little group:
\begin{eqnarray}
 a_1=a_2=a_3=1, a_4=...=a_8=0
\end{eqnarray}

In this case the space of spinors $Q$ decomposes into vector SO(3)
$q_i, i=1,2,3$ and 6 singlets $q_i, i=4,...,8$, which we represent
by zero operators, and it remains to represent relation

\begin{eqnarray}
 \{ q_i,q_j \}=  \delta_{ij}\\
 i,j=1,2,3. \nonumber
\end{eqnarray}

The minimal representation would be by Pauli's sigma matrices.
However, properties of that representation contradict to fermionic
nature of Q (and q) operators, since in that representation
product of all three q-s is 1. So, one can choose  double
dimensional (i.e. 4-dimensional) representation of q, in other
words, one can take three of four SO(4) gamma-matrices as $q_i$.
That will give a 4-dimensional representation by SO(4) spinors,
decomposition of which with respect to the SO(3) subgroup gives
two two-dimensional complex SO(3) spinors. The supercharges Q (q)
are transforming one spinor into another, so one of them should be
assigned Bose statistics, and Fermi statistics should be assigned
to the other one. This situation is similar to preon case, when
supermultiplet has two members, both in the same representation of
little group. The only difference is, that in the preon case
supercharges are not acting inside the representation of little
group. In the present case they do act, since 4-dimensional
gamma-matrices are both interchanging two spinors and acting
inside the spinor space. From the point of view of spin addition
rules, we have a spinor, on which a spin-1 supercharge is acting,
and in the final result the spinor representation is separated, by
the Pauli sigma-matrices, which is just an anti-diagonal entry of
4-dimensional gamma-matrix.

Beside the preon case, one can consider two other cases with SO(7)
little group. In one case we have first seven $a_i$ equal to 1
(e.g.) and remaining $a_8$ equal to zero, and in the other case
that last  $a_8$ is non-zero. In the first case we have little
group SO(7) acting on Q, $q_8$ is invariant, and set $q_i$ is
non-trivially transforming under SO(7). Evidently, it is
$\underline{7}$, vectorial, fundamental irrep of SO(7), since
there is no other representation(s) which can add up to give that
dimensionality. Again q-s anticommute as 7-dimensional
gamma-matrices, but we can't use the minimal 8-dimensional
representation, and should take bigger one, that of SO(8), by
16-dimensional gamma-matrices, i.e. the same representation as in
a particle case. Surprising, although simple fact is that $8_s$
and $8_v$ of SO(8) reduce, under spinorially embedded SO(7) into
SO(8) (\ref{so7}), to $\underline{7_v}+\underline{1}$ (vector plus
singlet) and $\underline{8_s}$ (Majorana spinor), respectively.
I.e. vector irrep of SO(8) gives rise to spinorial of SO(7), and
vice versa, spinor SO(8) decomposes into vector $\underline{7}$ of
SO(7) plus singlet. First statement is evident, and follows
immediately from \ref{so7}, second one can be checked directly,
and is evident, also, from the point of view of saturating 8 by
tensorial representation(s) of SO(7).

Finally, in the second case Q-s are represented in the general
form (\ref{Qq}), with the same SO(7) decomposition, and eighth
gamma-matrix plays a role of $q_8$.

In other cases the considerations are similar. Consider e.g. SO(4)
case (\ref{so4}). Representation will be exactly as of SO(3) since
in this last case we already used SO(4) matrices, and
decomposition of the representation space w.r.t. the SO(4) group
gives two spinor representations of SO(4) - Weyl and anti-Weyl.
Similarly, for SO(5) and SO(6) groups we use SO(6) gamma-matrices,
and decomposition of corresponding space of spinors, in which
these gamma-matrices are acting, gives two spinors, which have
opposite chiralities in the SO(6) case.

So, all these cases give the same phenomena - the supermultiplets
consist of two spinor representations of little group SO(k), to
one of which Fermi statistics should be assigned, and Bose
statistics to the other one.

\section{Rank Two Z Matrix}
For rank 2 we can present the complete results, including massive
case. General (positively defined) Z matrix in that case can be
presented in the form
$Z_{\alpha\beta}=\lambda_\alpha\lambda_\beta+\mu_\alpha\mu_\beta$
where $\lambda, \mu$ are 16-component Majorana-Weyl spinors. Since
space of non-zero spinors is homogeneous under Lorentz
transformation, we can take $\lambda$ as e.g. (in our
representation of gamma-matrices) $\lambda=(1,0,0,0,0,1,0,...)$.
This spinor is invariant w.r.t. the $T_8 \ltimes SO(7)$, so we can
use this remaining invariance to bring $\mu$ to some standard
form. $\mu$ can be decomposed w.r.t. the SO(8) into a couple of
8-dimensional spinors - Majorana-Weyl $\mu_1$ and
Majorana-anti-Weyl $\mu_2$. $T_8$ group is leaving invariant the
Weyl spinor and is transforming anti-Weyl spinor into Weyl one. If
anti-Weyl part, i.e. $\mu_2$, is zero, $\mu_1$ decomposes w.r.t.
the SO(7) (remember that it is spinorially embedded into SO(8))
into $ \underline {7} +\underline{1}$. $\underline{1}$ is
evidently proportional to (upper half of) $\lambda$. The
$\underline {7}$ can be brought by SO(7) to any form, except that
the scale ($\overline{\mu_1}\mu_1)$ cannot be changed, so one can
choose e.g.the following form:

\begin{equation}\label{m1}
  \mu_1=(a,b,0,0,-b,a,0,0)
\end{equation}
with real $a, b, b \neq 0$.

If $\mu_2$ is non-zero, then it (being spinor under SO(7)) can be
brought under SO(7) to any given form, e.g.

\begin{equation}\label{m2}
  \mu_2=(c,0,0,0,0,c,0,0)
\end{equation}
with real $c$, which leaves, by definition, the $G_2$ group of
invariance. This last group can be used to bring the $\mu_1$
spinor to some standard form, which, finally, can be brought to
zero by $T_8$ transformations. So, in this last case we have
nonzero $\mu_2$ (\ref{m2}), only.

The little group for the first case is actually calculated in
previous section, and for a general $a, b$ is a semidirect product
of $T_8$ to a compact group $SO(6)$, since there are 6 zero and
two different non-zero eigenvalues of corresponding Z matrix:

\begin{equation}\label{eig}
1+a^2+b^2\pm \sqrt{(1+a^2+b^2)^2-4b^2}
\end{equation}

Only for $a=0, b=1$ these eigenvalues coincide, and we have, in
agreement with above, the little group $T_8 \ltimes (SO(2) \otimes
SO(6))$.

For the second case the little group is $G_2 \otimes SO(2)$:

\begin{equation}\label{G2}
\begin{array}{*{20}c}
   0 & 0 & 0 & 0 & 0 & 0 & 0 & 0 & {\frac{{(1 - c)s_{89} }}{{1 + c^2 }}} & 0  \\
   . & 0 & {s_{12} } & {s_{13} } & {s_{14} } & {s_{15} } & {s_{16} } & {s_{17} } & 0 & 0  \\
   . & . & 0 & {s_{23} } & {s_{24} } & {s_{25} } & {s_{26} } & {s_{27} } & 0 & 0  \\
   . & . & . & 0 & {s_{34} } & {s_{35} } & {s_{36} } & {s_{26}  - s_{14} } & 0 & 0  \\
   . & . & . & . & 0 & { - s_{27}  - s_{36} } & {s_{35}  - s_{12} } & {s_{13}  + s_{25} } & 0 & 0  \\
   . & . & . & . & . & 0 & {s_{17}  - s_{34} } & { - s_{16}  - s_{24} } & 0 & 0  \\
   . & . & . & . & . & . & 0 & {s_{15}  - s_{23} } & 0 & 0  \\
   . & . & . & . & . & . & . & 0 & 0 & 0  \\
   . & . & . & . & . & . & . & . & 0 & {s_{89} }  \\
   . & . & . & . & . & . & . & . & . & 0  \\
\end{array}
\end{equation}

Eigenvalues of so(2) part of this matrix (i.e. sub-matrix at
intersection of 1-st, 9-th and 10-th rows and columns) are
\begin{equation}\label{so2}
  (0,i\frac{{4cs_{89} }}{{1 + c^2 }}, -i\frac{{4cs_{89} }}{{1 + c^2
  }})
\end{equation}
which confirms the compact SO(2) nature of corresponding
one-parameter group.

\section{Notations}
The space-time metric is  $\eta^{\mu\nu}$ = (-1,+1,...,+1). Gamma
matrices:
\begin{equation}\label{g}
\left\{\Gamma^{\mu}, \Gamma^{\nu}\right\} = \eta^{\mu\nu},
\mu,\nu=0,1,...,D-1
\end{equation}

are in the following representation. First we construct them in
Euclidean case, then for Minkowski metric one has to multiply
first matrix on imaginary unit $i$. We construct gamma-matrices by
induction. In 2 dimensions:

$\Gamma^0 = \left(%
\begin{array}{cc}
  0 & 1 \\
  1 & 0 \\
\end{array}%
\right) $, $
\Gamma^1= \left(%
\begin{array}{cc}
  0 & -i \\
  i & 0 \\
\end{array}%
\right) $

In dimension D = 2k+1 we use the matrices from D=2k, and define
one additional as a product of all 2k matrices, with appropriate
coefficient to ensure that it squares to 1:

$ (\Gamma)^{\mu}_{(D)\alpha\beta} =
(\Gamma)^{\mu}_{(D-1)\alpha\beta}$    ($\mu $ = 1,...,D-1)

 and

$(\Gamma)^{\mu}_{(D)} = \Gamma^{D}_{(D-1)}$ or
$(\Gamma)^{\mu}_{(D)} = i \Gamma^{D}_{(D-1)}$ ($\mu$ = D)

where we denote by $\Gamma^{D}_{(D-1)}$ the product of all gamma
matrices in dimension D-1=2k. And we choose one of those
conditions depending on which one of those satisfies Euclidean
Clifford algebra condition.

In dimension D = 2k we define

$\Gamma^{1}_{(D)} = \left(%
\begin{array}{cc}
  \bold{0} & -i \bold{1} \\
  i\bold{1} & \bold{0} \\
\end{array}%
\right)$

where $\bold{0}$ and $\bold{1}$ are zero and identity matrices of
proper dimensionality.

and

$\Gamma^{\mu}_{D} = \left(%
\begin{array}{cc}
  \bold{0} & \Gamma^{\mu-1}_{D-1} \\
  \Gamma^{\mu-1}_{D-1} & \bold{0} \\
\end{array}%
\right)$

It is easy to check that chosen $\Gamma$ matrices are indeed the
representation of Clifford algebra (\ref{g}).

Spinor representation of Lorentz algebra is  constructed by these
$\Gamma$ matrices, the generators of Lorentz transformations in
spinor space are the matrices $(1/4)[\Gamma^{\mu}, \Gamma^{\nu}]$.

\section{Acknowledgements}

We are indebted to R.Manvelyan for discussions. Work is partially
supported by INTAS grant 03-51-6346.

\end{document}